# Photon properties of single graphene nanoribbon microcavity laser

**Guangcun Shan**[1,2,*]

[1] *Department of Physics and Materials Science, City University of Hong Kong, Hong Kong SAR, China*

[2] *State Key Laboratory of Functional Materials for Informatics, Shanghai Institute of Microsystem and Information Technology, Shanghai, China,*

*Correspondence should be addressed to

Guangcun Shan

Department of Physics and Materials Science,

City University of Hong Kong, Hong Kong SAR, China

Phone: +852-34422279

Fax: +86-21-39966606

E-mail: guangcunshan@mail.sim.ac.cn

**ABSTRACT**

In this work, I propose a scheme about a single graphene nanoribbon (GNR) emitter in a microcavity, and focus on a fully-quantum-mechanical treatment model with the excitonic interaction included to investigate the photon properties and lasing action. When the single armchair-edged GNRs (AGNRs) microcavity system is pumped, the exciton-photon coupling provides more photons and enhances the photon emission process, making it essentially a lasing object. The theoretical results demonstrated that single AGNR in a semiconductor microcavity system maybe serve as a nanolaser with ultralow lasing threshold.

# 1. Introduction

In the past decade, considerable progress has been made in the fabrication of high-Q semiconductor microcavity with quasi-three-dimensional photon confinement and single nanostructure emitter[1-3]. In the context of quantum electrodynamics (QED) and single-photon source, the photon statistics and lasing properties of single nanostructure emitter interacting with the electromagnetic field of microcavity have received considerable interest[1-5]. Single nanostructure-based microcavity laser, in particular, the single quantum dot (QD) lasers proposed as the solid-state analog to the one-atom laser have been investigated both for fundamental aspects [1-3] and for applications [4,5]. On the other hand, since the successful fabrication and measurement of graphene[6,7], there have been intensive research on graphene-based nanostructures because of fundamental physics interests and promising applications[8]. Furthermore, a strip of graphenes e wide strip of referred to as graphene nanoribbons (GNRs) can be obtained by etching or patterning[9]. Recent advances in fabricating and characterizing stable GNRs[9-15] have opened the way to explore their various remarkable properties of these systems. In 2007, Cohen's studies[11,15] have shown that the quasi-particle band gap of armchair-edged GNRs (AGNRs) is in the interesting energy range of 1-3 eV for 2-1 nm wide GNRs due to many-electron effects and is therefore promising for optoelectronics applications. Since then, they were immediately recognized as one of the building blocks for nanophotonics. As a result, by placing a single AGNR in a high finesse external microcavity or fabricating a distributed Bragg reflector with high reflectivity, it is possible to achieve a single AGNR microcavity laser with ultralow lasing threshold.

In light of these issues above, in this work, I propose the single AGNR microcavity schematic structure and model, and examine the sense in which single AGNR in a microcavity can be regarded as a nanolaser. The outline of the paper is as follows. In the next section, I present the schematic structure of the single GNR microcavity system as well as the theoretical model with the excitonic interaction included in the Lindblad form. Section 3 shows the results about the photon emission of a single GNR in a microcavity system in both the strong-coupling regime and the weak coupling regime. Finally, in section 4, a conclusion of the results is drawn.

## 2. Theory and Model

As a specific scheme illustration, the considered system is a single AGNR embedded into a semiconductor microcavity. The internal two sides of the cavity mirrors are separated by a distance which is of the multiple integer of the lasing wavelength. Inside the microcavity there is a single AGNR localized in a position which corresponds to most of the electromagnetic field.

Since single-nanostructure-emitter laser, such as one-atom laser, single QD microcavity laser, etc., may operate with only a small number of photons in the lasing microcavity, quantum-mechnical processes may be occurring and a fully quantum-mechnical treatment including the quantization of the field must be used. In an atomic model for single QD microcavity system, one usually assumes lasing transition between two electronic configrations from an excited 1S state to a lower ground state. In the corresponding theoretical models, this coupling is described the Jaynes-Cummings Hamiltonian, and thereby such a single QD microcavity laser system have been mainly modeled with atomic models[4] that usually don't include the excitonic interaction effect. Yet, for single AGNR in a microcavity, the analogy breaks down in several important respects, due to the reduced dimensionality effects and special quasi-particle band gap of AGNRs. To a good accuracy, in AGNRs the conduction and valence band states have pairwise equal wave functions and an excitonic carrier jump between lasing states is different from the atomic QDs[15]. The electromagnetic field can excite an electron of the valence band to the conduction band by creating a hole in the valence band. The electron and the hole interact by giving excitonic states [11,14,15]. The lasing transition is from an excited state to a lower ground state, and the excitonic state 1S which is the lowest bound state has the main oscillater strength[1,17]. For this reason, we could take into account only this state for the exciton-photon interaction. As a consequence, for a single AGNR with the strong quantum confinement effects, we should take excitonic interaction into account for the single AGNR in a microcavity system because of reduced dimensionality.

In the dipole approximation, the effective linear Hamiltonian $H$ for the exciton-photon coupling system in the cavity is $H = \sum_{\mathbf{k}} H_{\mathbf{k}}$ with

$$H_{\mathbf{k}} = \hbar\omega_{\mathbf{k}c} a_{\mathbf{k}}^+ a_{\mathbf{k}} + \hbar\omega_{\mathbf{k}exc} b_{\mathbf{k}}^+ b_{\mathbf{k}} + i\hbar g\left(a_{\mathbf{k}}^+ b_{\mathbf{k}} - b_{\mathbf{k}}^+ a_{\mathbf{k}}\right) + \sum_{q} \hbar\alpha b_{\mathbf{k}-q}^+ b_{\mathbf{k}+q}^+ b_{\mathbf{k}} b_{\mathbf{k}},  \qquad (1)$$

where the photon creation and annihilation operators for photons of electromagnetic fields of microcavity with a wave vector $\mathbf{k}$ are $a_{\mathbf{k}}^+$ and $a_{\mathbf{k}}$, respectively; and the exciton creation and annihilation operators with a wave vector $\mathbf{k}$ are $b_{\mathbf{k}}^+$ and $b_{\mathbf{k}}$, respectively; $\omega_{\mathbf{k}exc}$ is the transition frequency of the excitonic mode with a wave vector $\mathbf{k}$; $\omega_{\mathbf{k}c}$ is the cavity resonant frequency with a wave vector $\mathbf{k}$, and the coupling constant of the exciton-photon interaction is thus given by $g = \left(\dfrac{1}{4\pi\varepsilon_0\varepsilon_r}\dfrac{e^2 f}{mV}\right)^{1/2}$ [1-3,16-19], where $f$ is the exciton oscillator strength, $V$ the effective modal volume, $m$ the free electron mass, $e$ the electron charge, and $\varepsilon_0\varepsilon_r$ the dielectric constant; The parameter $\alpha$, which represents the strength of the excitonic interaction, has the following expression [17,18]

$$\alpha = \frac{3a_B E_{ex}}{S}, \qquad (2)$$

where $S$ is the quantization area; $a_B$ and $E_{ex}$ represent respectively excitonic Bohr radius and binding energy. The fourth term describes the effective exciton–exciton scattering due to Coulomb interaction. It should be noted that the value of the nonlinear excitonic parameter which describes the strength of the interaction between excitons can be estimated to be $\alpha=0.01$ and the details can be found in Ref. [16-18,22].

In the present case of single AGNR in a microcavity with a resonant excitation at normal excitation incidence pump (only $\mathbf{k}=0$ direction), for the effective linear Hamiltonian $H$, we have $H \cong H_{\mathbf{k}=0}$. Therefore, in the following, the AGNR can be described as a two-level system with the ground state $|0\rangle$ and the singlet excitonic state $|E\rangle$, where $H_0|0\rangle = E_0|0\rangle$ for the ground state, and $H_0|E\rangle = E_1|E\rangle$ for the singlet excitonic state. The Hamiltonian is further given by

$$H_0 = \hbar\omega_c a^+ a + \hbar\omega_{exc} b^+ b + i\hbar g\left(a^+ b - b^+ a\right) + \hbar\alpha b^+ b^+ bb \qquad (3)$$

Now I'll examine the sense in which single AGNR in a microcavity can be regarded as a nanolaser and work out the possible lasing effects. Given sufficiently strong exciton-photon coupling interaction between the AGNR and the electromagnetic field in the microcavity, light emission is either promoted when it resonates with a cavity eigenmode, or suppressed when it is off-resonant. The photonic and excitonic modes are quantified along the normal direction to the microcavity. Invariance by translation implies that the excitons with parallel wave vector $K_{//}$ can couple only with photons with parallel wave vectors $k_{//} = K_{//}$. This work deals only with the case of normal incident pumping mode irradiating the microcavity. Here the cavity reduces the spontaneous emission lifetime of the radiative transition, increasing the spontaneous emission rate. Here we've taken the excitonic spontaneous decay rate $\gamma=1$ ns$^{-1}$ as a characteristic constant. The schematic energy level and exciton-photon coupling of single AGNR in a microcavity system is illustrated in Fig.1.

Applying the standard methods of the quantum theory of damping, the master equation in the general Lindblad form can be governed by in the rotating frame [17-23],

$$\frac{\partial}{\partial t}\rho = L\rho = \frac{1}{i\hbar}[H,\rho] + L_{diss}\rho + L_P\rho, \tag{4}$$

where, $L_{diss}$ is the dissipation terms. The dissipations lead to nonunitary evolution given by [17-21]

$$L_{diss}\rho = -\frac{\gamma}{2}\left(b^+b\rho + \rho b^+b - 2b\rho b^+\right) - \kappa\left(a^+a\rho + \rho a^+a - 2a\rho a^+\right), \tag{5}$$

$$L_P\rho = -\frac{\Gamma_L}{2}\left(bb^+\rho + \rho bb^+ - 2b^+\rho b\right) \tag{6}$$

where the excitonic spontaneous emission with the rate $\gamma$ as well as the cavity decay with the rate $\kappa$ are included, , and $\Gamma_L$ is the incoherent excitation pumping rate of the GNR. Specifically, the master equation (4) can be written as a first-order differential equations using the truncated basis set of joint number (photon)-exciton states $\{|n,\alpha\rangle, n=0,1,2,\ldots,N; \beta=0,E\}$, where $n$ denotes the photon number state, and *α* denotes the exciton state of AGNR. The master differential Equation (4) is solved using the Savage and Carmichael simulation method [18-21].

## 3. Results and discussions

Firstly, we investigate the dynamics of single GNR in a microcavity system. The mean photon number $\langle n \rangle$ versus time evolution for different values of coupling strength *g* and pump $\Gamma_L$ is shown in Fig. 2. As indicated in Fig. 2, increasing *g* or $\Gamma_L$ decreases the time for the mean intracavity photon number to reach the steady-state. In order to achieve the possibility of lasing action for this single AGNR microcavity system, one needs to obtain the population inversion. For the fully-quantum mechanical treatment case, the inversion is given by

$$\Delta_{inversion} = \sum_{n=0}^{\infty} \left( \langle n, E | \rho | n, E \rangle - \langle n, 0 | \rho | n, 0 \rangle \right) \tag{7}$$

This happens when sufficient population inversion, and consequently net stimulated photon emission is achieved. Furthermore, mean photon number $\langle n \rangle$ in a microcavity versus pump strength $\Gamma_L$ for different coupling strength *g* is shown in Fig. 3. Here, we find that $\langle n \rangle$ increases as the pump strength $\Gamma_L$ is increased. The reason is that the pump increases the coherence between the lasing levels. This property behaves similarly to the one-atom laser [18,19,23]. It is seen that decreasing *κ* increases the intracavity photon number, which also behaves similarly to the one-atom laser [18,19,21,23]. From the simulation results, experimentally speaking, we infer that it may be possible to achieve a GNR microcavity laser if *g*>4*γ* and the cavity is a suitable cavity with *κ*≤ *γ*. Under these conditions, the photon bottleneck is essentially broken up by resonance emission due to the coupling interaction. Moreover, the cavity field leads to much more photons than alone via excitonic spontaneous decay of AGNR, making the AGNR into a lasing emitter essentially in analog to one-atom laser or single QD microcavity laser[2-4,19]. We also find that it has something in common with microlasers, and a reduction of spontaneous emission into the lasing mode.

Here, since only interested in the steady state properties, the semiclassical factorization of the density matrix is used for larger photon numbers. Semiclassically a laser is said to be at threshold when the pumping rate is just sufficient for photon production to balance the losses. As a result, it is possible to derive a semiclassical approximation for the mean photon number and the lasing inversion in steady state. Moreover, the expectation values of exciton and field operators are assumed to factorize, since the photon number in the present AGNR microcavity system is large enough. Solving the equations above, we find above the threshold

$$\langle n \rangle = \frac{1}{\kappa}(\Gamma_L - \gamma) - \frac{\kappa}{2g^2}(\gamma + \Gamma_L) \qquad (8)$$

And the laser threshold pump rate is $\Gamma_{threshold} = \frac{g^2}{\kappa} - \gamma$. Finally, the population inversion is given by

$$\Delta_{inversion} = \begin{cases} \kappa(\gamma + \Gamma_L)/2g^2 & above \quad threshold \\ -1/\Gamma_L & below \quad threshold \end{cases} \qquad (9)$$

Laser threshold is reached if the mean number of photons in the lasing mode is more than one. At this critical point, stimulated emission overtakes spontaneous emission, and linear amplification is replaced by nonlinear laser oscillation. As indicated by Eqs. (8) and (9), since the excitonic parameter α is absent from both of the equations, the laser threshold is independent of α, and laser threshold can be reached for all $\kappa$ values considered. This means the single AGNR in a microcavity can serves as a nanolaser with ultralow threshold. Here, it is worth pointing out that the lasing photon spectrum should be dependent on the excitonic parameter α.

## 4. Conclusion

In summary, I have proposed a lasing model scheme about a single GNR in a microcavity with the excitonic interactions included and investigated the photon properties and lasing action. When the single AGNR microcavity system is pumped, the exciton-photon coupling provides more photons and enhances the photon emission process, making it essentially a lasing object. The theoretical results demonstrated that single AGNR in a semiconductor microcavity system maybe serve as a nanolaser with ultralow lasing threshold.

**Acknowledgement**

I would like to thank Prof. Marvin Cohen for his stimulating discussion on exciton and optical properties of graphene nanoribbons during his visit to Institute for Advanced Study (IAS), Hong Kong on this October.

**References**

[1] Reithmaier, J. P., Sek, G., Loeffler, A., Hofmann, C., Kuhn S., Reitzenstein, S., Keldysh, L. V., Kulakovskii, V. D., Reinecke, T. L., Forchel, A. Strong coupling in a single quantum dot-semiconductor microcavity system. Nature **2004**, 432, 197–200.

[2] Peter, E.; Senellart, P.; Martrou, D.; Lemaître, A.; Hours, J.; Gérard, J. M.; Bloch, J. Exciton-photon strong-coupling regime for a single quantum dot embedded in a microcavity. *Phys. Rev. Lett.* **2005**, 95, 067401-1-4.

[3] Yoshie T.; Scherer A., Hendrickson J., Khitrova G., Gibbs H. M., Rupper G., Ell C., Shchekin O. B., Deppe D. G. Vacuum Rabi splitting with a single quantum dot in a photonic crystal nanocavity. *Nature* **2004**, 432, 200–202.

[4] Pelton M.; Yamamoto Y. Ultralow threshold laser using a single quantum dot and a microsphere cavity. *Phys. Rev. A*, **1999**, 59, 2418–2422.

[5] Nomura, M.; Iwamoto, S. Localized excitation of InGaAs quantum dots by utilizing a photonic crystal nanocavity. *Appl. Phys. Lett.* **2006**, 88, 141108-1-3.

[6] Novoselov, K. S.; Geim, A. K.; Morozov, S. V.; Jiang, D.; Zhang, Y.; Dubonos, S. V.; Grigorieva, I. V.; Firsov, A. A. Electric Field Effect in Atomically Thin Carbon Films. *Science* **2004**, 306, 666-669.

[7] Novoselov, K. S.; Geim, A. K.; Morozov, S. V.; Jiang, D.; Katsnelson, M. I.; Grigorieva, I. V.; Dubonos, S. V.; Firsov, A. A. Two-dimensional gas of massless Dirac fermions in grapheme*, Nature* **2005**, 438, 197-200.

[8] Geim, A. K.; Novoselov, K. S. The rise of graphene. Nat. Mater. **2007**, 6, 183-191.

[9] Chen, Z.; Lin, Y.-M.; Rooks, M. J.; Avouris, P. Graphene nano-ribbon electronics. *Physica (*Amsterdam*)* **2007**, 40E, 228-232.

[10] Li, X.; Wang, X.; Zhang, L.; Lee, S.; Dai, H. Chemically Derived, Ultrasmooth Graphene Nanoribbon Semiconductors. *Science* **2008**, 319, 1229-1232.

[11] Son, Y.-W.; Cohen, M. L.; Louie, S. G. Half-metallic graphene nanoribbons. *Nature* **2006**, 444, 347-349.

[12] Brey, L.; and Fertig, H. A. Electronic states of graphene nanoribbons studied with the Dirac equation. *Phys. Rev. B* **2006**, 73, 235411-1-5.

[13] Yang, L.; Park, C.-H.; Son, Y.-W.; Cohen, M.L.; Louie S.G. Quasiparticle energies and band gaps of graphene nanoribbons. *Phys. Rev. Lett.* **2007**, 99, 186801-1-4.

[14] Li, Y.; Xiaowei Jiang, X.; Liu, Z.; Liu, Z. Strain effects in graphene and graphene nanoribbons: The underlying mechanism. *Nano Res.* **2010**, 3, 545-556.

[15] Yang L., Cohen M.L., and Louie S.G. Excitonic effects in the optical spectra of graphene nanoribbons. *Nano Lett.* **2007**, 7, 3112-3115.

[16] Baas A., Karr J., Eleuch H., Giacobino E. Optical bistability in semiconductor microcavities. *Phys. Rev. A,* **2004**, 69, 023809-1-8.

[17] Eleuch H. Photon statistics of light in semiconductor microcavities. *J. Phys. B: Atom., Mole. and Opt. Phys.* **2008**,41, 055502-055507.

[18] Shan, G.C.; Huang, W. Photon properties of light in semiconductor microcavities. *Front. Optoelectr. in China*, **2009**, 2, 345-349.

[19] Mu, Y.; Savage, C. M. One-atom lasers. *Phys. Rev. A* **1992**,46, 5944-5954.

[20] Savage, C. M.; Carmichael, H. J. Single atom optical bistability. *IEEE J. Quantum Electron.* **1998**, 24, 1495-1498.

[21] Scully, M. O.; Zubairy, M. S. Quantum optics. London: Cambridge University Press, 190-220 (1997).

[22] Eastham, P. R.; Littelewood, P. B. Finite-size fluctuations and photon statistics near the polariton condensation transition in a single-mode microcavity. *Phys. Rev. B*, **2006**, 73, 085306-01-11.

[23] Carmichael, H.; Luis, A. O. Quantum optics: single atom lases orderly light. *Nature* **2003**, 425, 246-247.

**Figure Captions**

Fig. 1. Schematic energy level and exciton-photon coupling of single AGNR in a microcavity.

Fig.2. Mean photon number <n> versus time evolution for various values of $\kappa$ and $\hbar g$ (from the bottom to the top) of 80μeV(red), 400μeV (green), and 500μeV (blue), respectively, with $\gamma= 1$ ns$^{-1}$, $\Gamma_L =10$ ns$^{-1}$, and $\kappa = 0.1$ ns$^{-1}$.

Fig.3. Mean photon number <n> versus pump rate $\Gamma_L$ at various exciton-photon coupling strength $\hbar g$ (from the bottom to the top) of 10μeV (red), 50μeV (green), 80μeV (blue), 160μeV (yellow), and 400μeV (black), respectively, with $\gamma= 1$ ns$^{-1}$. (a) $\kappa = 1$ ns$^{-1}$. (b) $\kappa = 0.1$ ns$^{-1}$.

**Figure 1.**

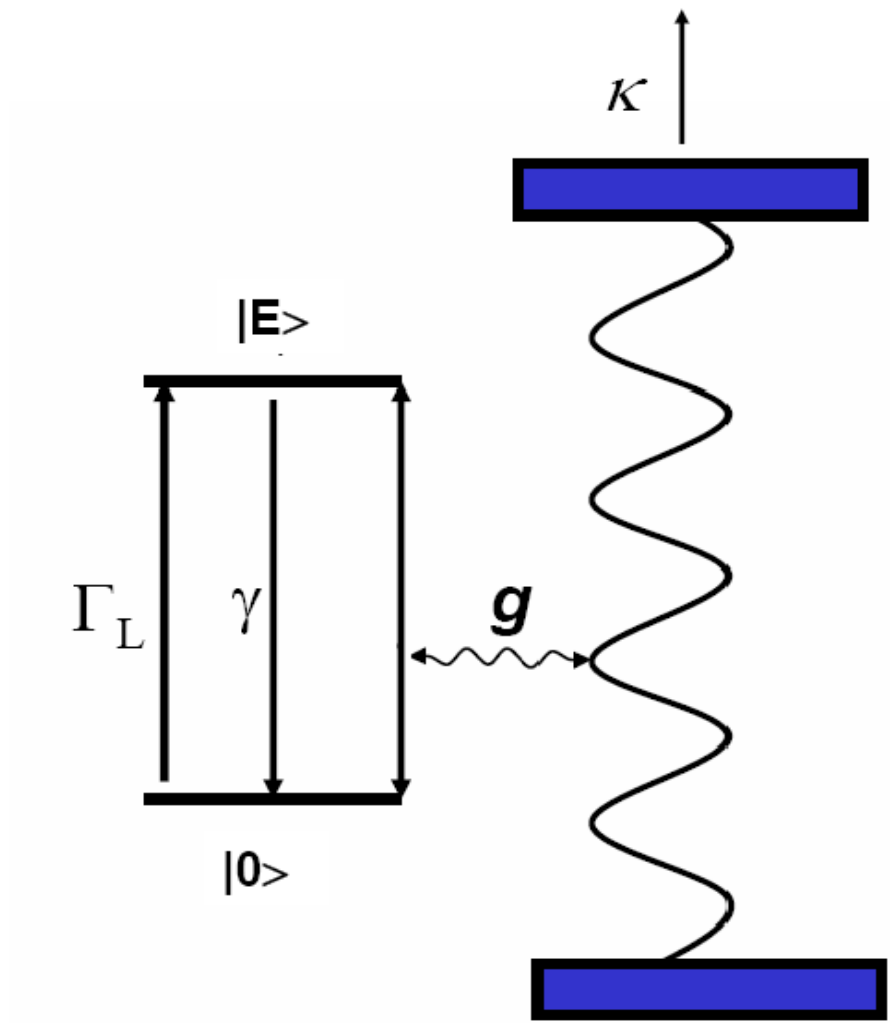

**Figure 2.**

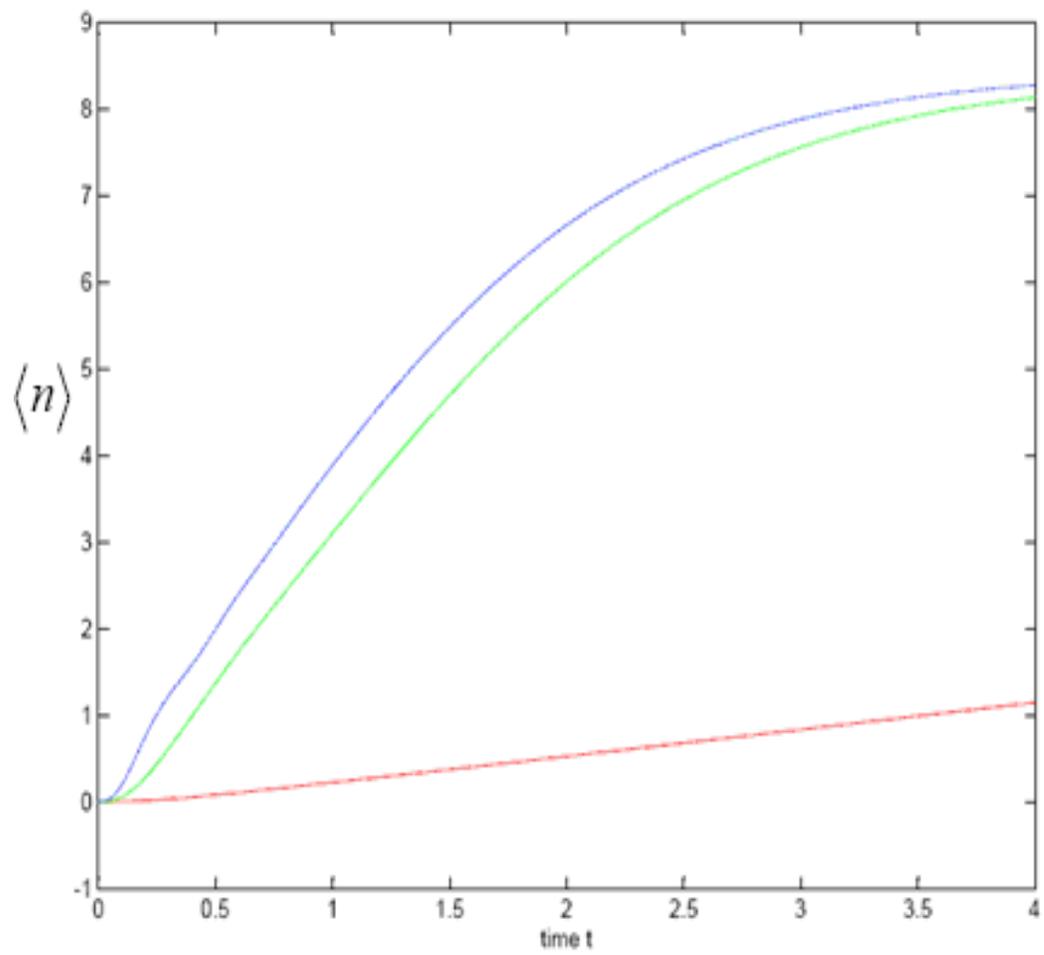

Fig.3.

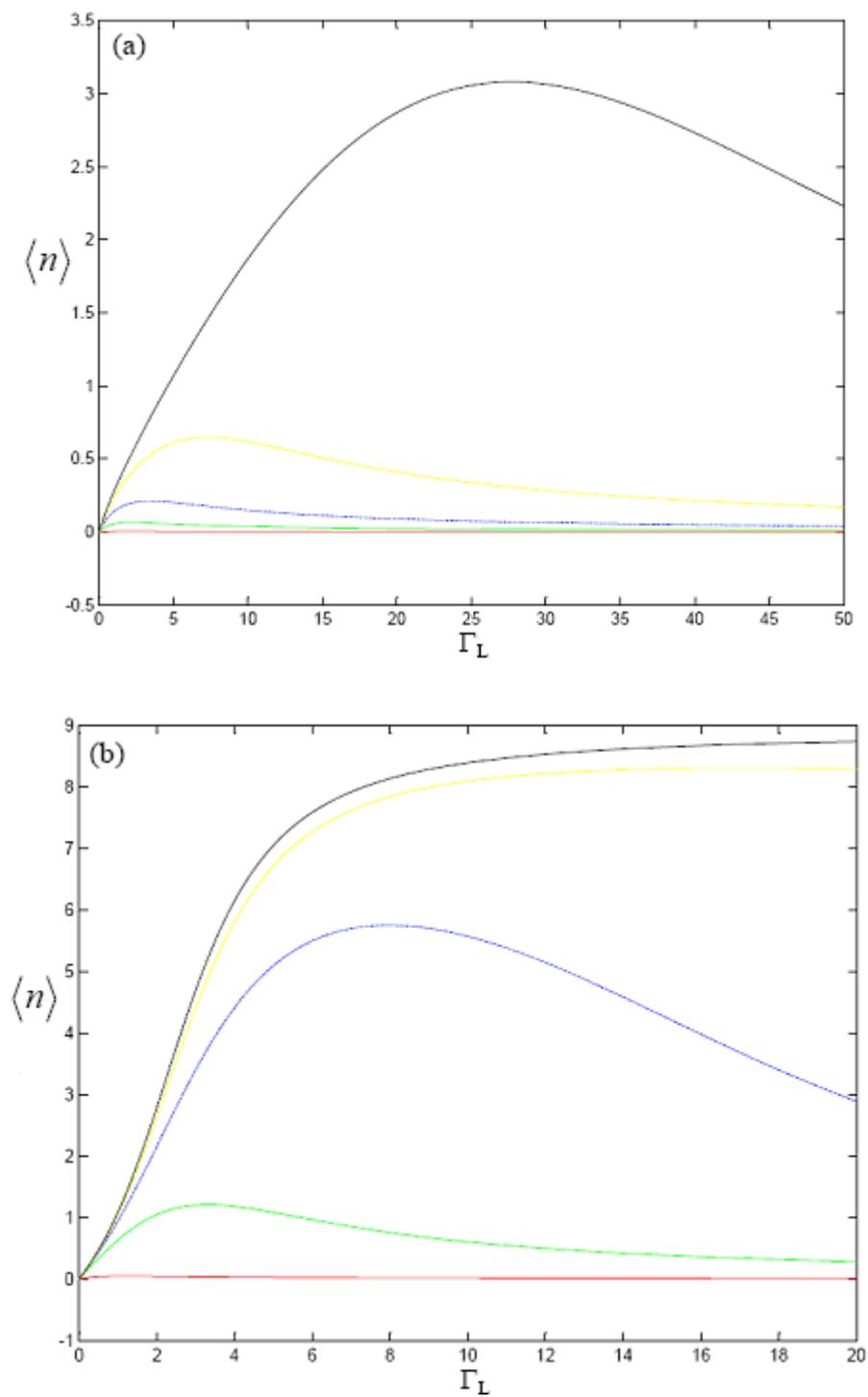